\documentclass[twocolumn]{aastex701}

\usepackage{svg}
\usepackage{xcolor}

\begin{document}

\title{Neuro-Parametric Spectral Classification of \\ Black Hole and Neutron Star X-ray Binary Systems}

\author{Akash Garg}
\thanks{Equal Contribution}
\affiliation{Inter-University Centre for Astronomy and Astrophysics (IUCAA), Pune, Maharashtra 411007, India}
\affiliation{Persistent Systems Ltd., Pune, Maharashtra 411006, India}
\email[show]{akash.garg@iucaa.in}

\author{Aman Kumar}
\thanks{Equal Contribution}
\affiliation{Inter-University Centre for Astronomy and Astrophysics (IUCAA), Pune, Maharashtra 411007, India}
\affiliation{Department of Physics, Savitribai Phule Pune University, Ganeshkhind, 411007, Pune, India}
\email[show]{kumaramanasci@gmail.com}

\author{Ajit Kembhavi}
\affiliation{Inter-University Centre for Astronomy and Astrophysics (IUCAA), Pune, Maharashtra 411007, India}
\email{akk@iucaa.in}
\author{Ranjeev Misra}
\affiliation{Inter-University Centre for Astronomy and Astrophysics (IUCAA), Pune, Maharashtra 411007, India}
\email{rmisra@iucaa.in}
\author{Aniruddha Kembhavi}
\affiliation{Wayve AI, London, UK}
\affiliation{University of Washington, Seattle, WA, USA}
\email{anikem@gmail.com}
\author{N. S. Philip}
\affiliation{Inter-University Centre for Astronomy and Astrophysics (IUCAA), Pune, Maharashtra 411007, India}
\affiliation{Artificial Intelligence Research and Intelligent Systems, Kerala, India}
\email{ninansajeethphilip@gmail.com}
\author{Rohan Pattnaik}
\affiliation{William H. Miller III Department of Physics \& Astronomy, Johns Hopkins University, MD, USA}
\email{rpattna1@jh.edu}
\author{Shreya Watwe}
\affiliation{MKSSS's Cummins College of Engineering for Women, Pune, India}
\email{watweshreya@gmail.com}
\begin{abstract}

We perform the classification of black hole and neutron star X-ray binary systems using deep neural networks applied to archival RXTE X-ray spectral data. We first construct two neural network models: one trained using only spectral flux values and another trained using both fluxes and their associated errors. Both models achieve high classification accuracies of  $\sim$90-94 \%. To gain physical interpretability of these networks, we fit all spectra with a simple phenomenological model consisting of a thermal disk component and a power-law. From this analysis, we identify the blackbody temperature, power-law index, the ratio of blackbody to power-law flux, the reduced $\chi^2$, and the variance of the data as key parameters that likely contribute to the classification. We validate this inference by designing an additional neural network trained exclusively on this reduced parameter set, without using the spectral data directly. This parameter-based model achieves a classification accuracy comparable to that of the spectral models. Our results show that deep neural networks can not only classify compact objects in X-ray binaries with high accuracy but can also be interpreted in terms of physically meaningful spectral parameters derived from conventional X-ray spectral analysis. This framework offers a promising, mission-agnostic approach for compact object classification in current and future X-ray surveys.

\end{abstract}



\section{Introduction} 
X-ray binary (XRB) systems consist of a compact object and a companion star orbiting their common center of mass. The compact object, formed as the end product of stellar evolution, could be a black hole (BH), a neutron star (NS), or a white dwarf (WD), whereas the companion can range from a low-mass main-sequence star to a massive OB-type star. The companion star may undergo Roche lobe overflow or stellar wind flow, leading to matter transfer via the Lagrangian point (L1) towards the compact star. The angular momentum of the infalling matter is redistributed as it falls into the strong gravity of the compact object, leading to the formation of a planar accretion disk around the compact object \citep{frank85}. The accretion disk radiates across the electromagnetic spectrum, ranging from optical/UV in its outer regions to hard X-rays from its innermost regions, making XRBs key targets for several space-based observatories. \citep[e.g.][]{belloni10}

Through decades of observations, a large population of Galactic and extragalactic XRBs has been discovered \citep[e.g.][]{liu07,santana16,avakyan23,fortin24}. The XRBs are commonly classified according to (i) the nature of the compact object (BH-NS-WD XRBs), (ii) the mass of the companion star (low mass or high mass XRBs), and (iii) the variability in their X-ray luminosities (persistent or transient XRBs). A common approach to determining the class of XRB involves studying its spectral and temporal properties to infer the nature of the compact object and its companion \citep[e.g.][]{bahramian23}. Specifically, timing and spectral products, such as light curves, power spectra, and photon spectra, are modeled to estimate binary parameters and to probe accretion–ejection properties, which together help reveal the physical nature of the system. 

\begin{figure*}
    \centering
    \includegraphics[width=19.0cm, height=8cm]{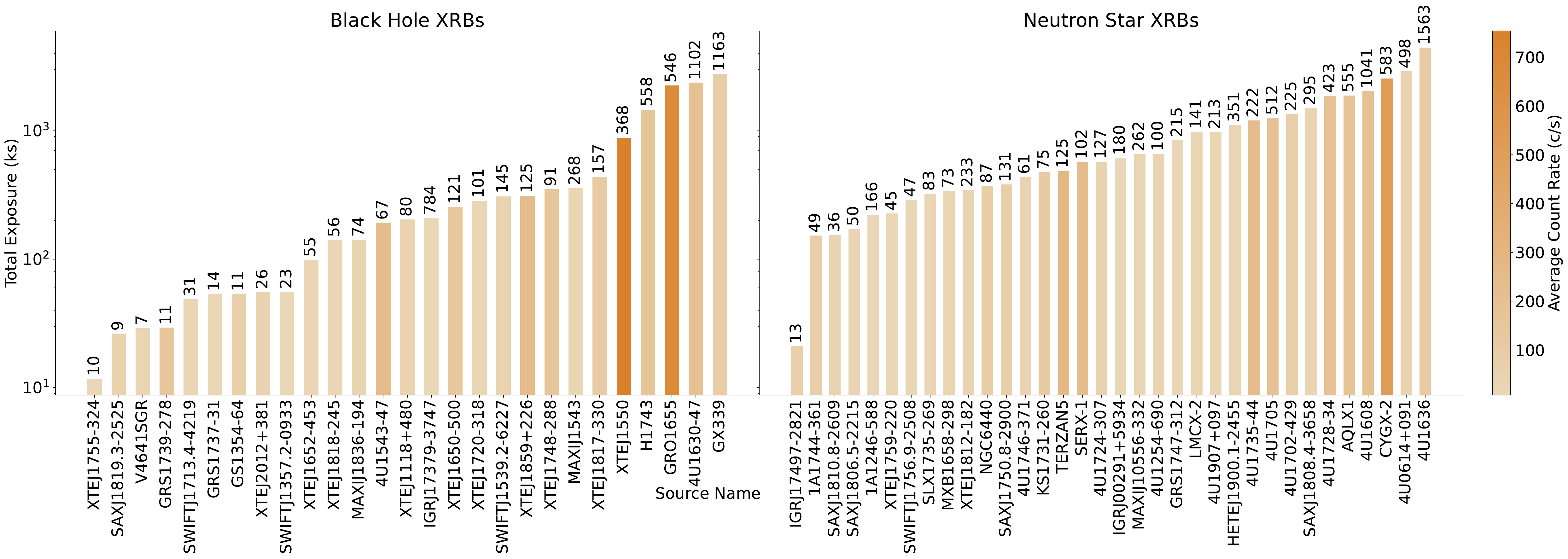}
    \caption{Distribution of RXTE observational coverage for the X-ray binary sample analysed in this work. The left panel shows black hole X-ray binaries (BH XRBs), while the right panel shows neutron star X-ray binaries (NS XRBs). For each source, the bar height indicates the total accumulated RXTE exposure time in kiloseconds (left y-axis, shown on a logarithmic scale), while the bar colour encodes the average source count rate in counts/sec, as indicated by the colour bar. The total number of individual RXTE observations available for each source is annotated above the corresponding bar. Sources are ordered by increasing total exposure within each class, highlighting the wide dynamic range in both exposure time and source brightness across the sample.}
    \label{fig:observation}
\end{figure*}

For instance, the time-averaged X-ray spectra of BH-XRBs typically show the presence of two dominant emission components in the accretion flow, namely, a low-energy thermal blackbody disk emission, peaking at a few tenths to a few keV, and a high-energy non-thermal power-law emission with an exponential cutoff at tens to a few hundred keV. These components are generally associated with an outer geometrically thin-optically thick accretion disk and an inner hot geometrically thick-optically thin coronal flow \citep[e.g.][]{gilfanov10}. In addition, a narrow emission feature at $\sim$ 6.5 keV may sometimes appear in the photon spectra, commonly interpreted as iron K$_\alpha$ fluorescence arising from reprocessing of the hard coronal emission by the accretion disk. NS-XRBs often display an additional low-energy blackbody component, attributed to emission from the NS surface or boundary layer, which provides an important observational signature to make its distinction from BH-XRBs possible \citep[e.g.][]{belloni18}. But this component may not always be present, and proper modelling of X-ray spectra is required to determine the nature of the compact object. Even then, spectral modelling alone is sometimes insufficient and must be supplemented by other diagnostics, such as the detection of thermonuclear bursts or coherent pulsations to establish the presence of a hard surface in NS X-ray binaries, or radial velocity measurements from optical or near-infrared spectroscopy to determine the mass function of the compact object (with a BH inferred if the mass $\geq 3 M_{\odot}$) \citep{bahramian23}.

\begin{figure*}
    \centering
    \includegraphics[width=0.48\linewidth]{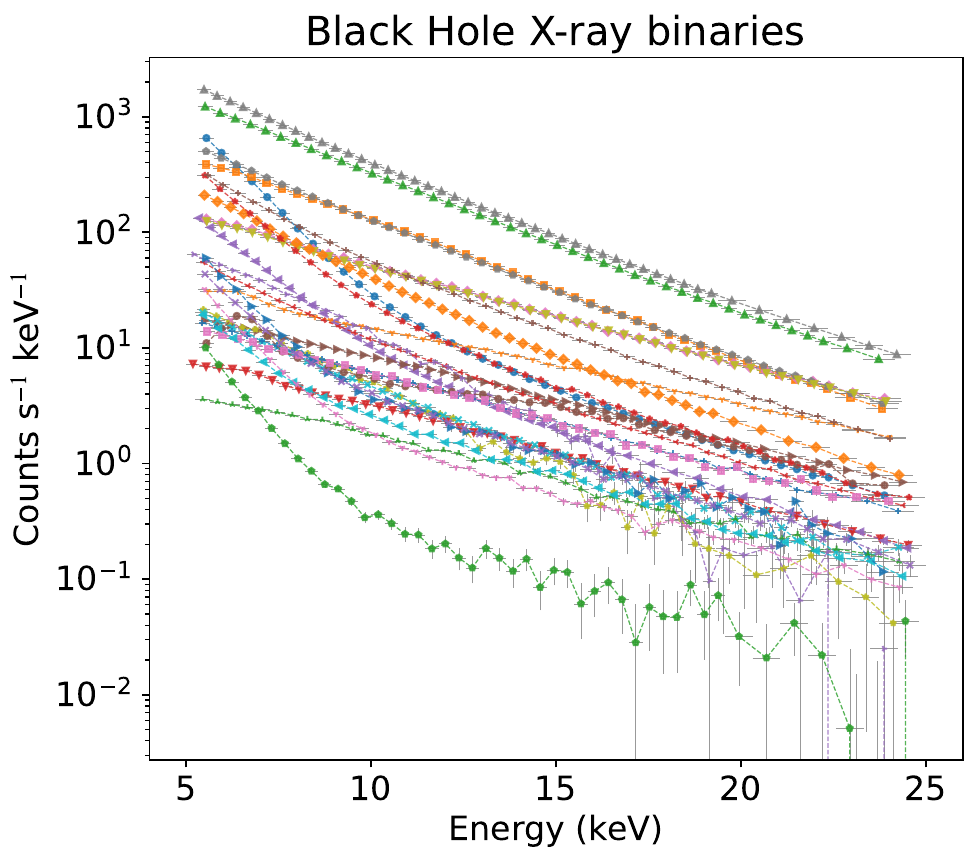}
    \includegraphics[width=0.48\linewidth]{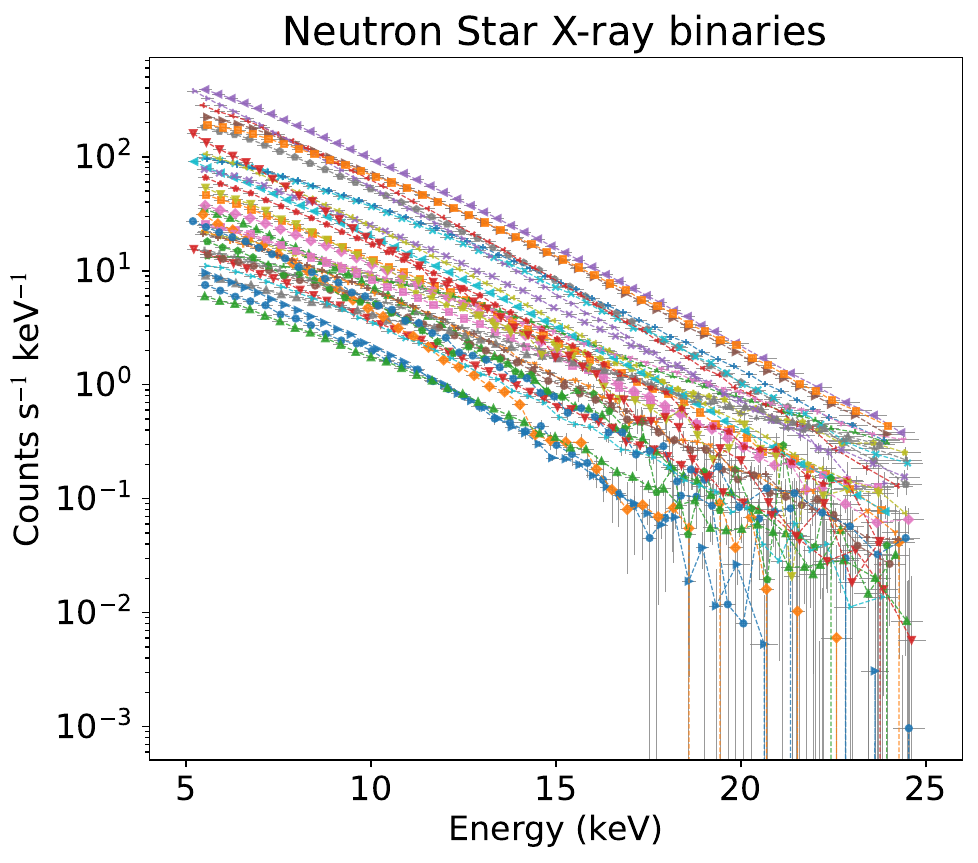}
    \caption{Representative RXTE energy spectra for the X-ray binary sample analysed in this work. The left panel shows black hole X-ray binaries (BH XRBs), while the right panel shows neutron star X-ray binaries (NS XRBs). For each source, we display the spectrum corresponding to the observation with the highest average count rate, chosen to maximise signal-to-noise. The spectra are shown in units ($\text{Counts~} \text{s}^{-1} \text{keV}^{-1}$) as a function of energy in the 5–25 keV range, with statistical uncertainties indicated by error bars. Each coloured curve corresponds to a different source, illustrating the diversity in spectral shapes and flux normalisations within and between the two populations.}
    \label{fig:spectra}
\end{figure*}

Traditional diagnostics, such as the one described above, continue to contribute to our better understanding of XRBs. However, over the past three decades, several dedicated X-ray telescopes such as the Rossi X-ray Timing Explorer (RXTE), Chandra, XMM-Newton, Neil Gehrels Swift Observatory, AstroSat, Nuclear Spectroscopic Telescope Array (NuSTAR), Neutron star Interior Composition Explorer (NICER), and Hard X-ray Modulation Telescope (HXMT) have been launched. These missions have operated over long durations, detecting a large population of XRBs and building extensive archives of observations. Additionally, recently launched wide-field telescopes such as the extended ROentgen Survey with an Imaging Telescope Array (eROSITA), Space Variable Objects Monitor (SVOM), Einstein probe, and X-Ray Imaging and Spectroscopy Mission (XRISM) are already detecting, or are expected to detect, millions of X-ray sources \citep[and hundreds of XRBs][]{fortin24}. Therefore, it requires modern techniques such as Artificial Intelligence (AI) and its subsets, Machine Learning (ML) and Deep Learning (DL), to efficiently handle the growth in data volume, analyze the observations, and obtain meaningful insights about the XRBs.

AI/ML has proven to be highly effective in performing downstream tasks in astrophysics, such as classification, regression, anomaly detection, and parameter inference on large astronomical datasets, thereby extending our knowledge of astrophysical sources and their properties \citep{kembhavi2022machine, sen2022astronomical, baron2019machine, barchi2020machine, cavuoti2015photometric}. The core idea of these learning methods is to construct a model/function that can best describe the underlying patterns in the data and help in making inferences about the system. The model parameters are optimized using training data, and then the model is further evaluated on an independent test dataset \citep{mitchell1997}. Learning can be 'supervised' if the data is labeled beforehand, providing an explicit input-output mapping during training. For instance, this approach is particularly useful for tasks such as binary classification between two types of sources (as in this work) by constructing a trained classifier model to determine the nature of the source. However, it is often challenging to label a large amount of data for training, and in such cases, 'unsupervised' methods are used to uncover hidden patterns in unlabeled data as shown by \citep{hayat2021self,huertas2023brief, bhatta2024gamma}. These methods are widely used in astronomy to reveal new classes of objects or to identify previously unknown members of known classes. DL, which is a subfield of ML, extends these approaches by using multi-layered neural networks that can automatically learn highly complex patterns from the data \citep{lecun15}. This makes DL especially powerful when dealing with large, high-dimensional astronomical datasets (further details are provided in the next section).

Numerous studies have employed supervised and unsupervised machine learning and deep learning techniques in X-ray astronomy. For instance, \cite{daniela17} used a supervised random forest to classify the spectro-temporal states of the BH-XRB GRS 1915+105 using its RXTE observations. Similarly, \cite{orwat22} applied a long-short-term memory variational autoencoder with a Gaussian mixture model on RXTE light curves of GRS 1915+105. \cite{beurs22} classified BH and NS-XRBs using 3D color-color-intensity diagrams with Bayesian Gaussian Processes, K-Nearest Neighbours, and Support Vector Machines. \cite{kumaran23} went a step ahead and applied the LGBM (Light Gradient Boosted Machine) algorithm to classify Chandra point sources into Active Galactic Nuclei (AGN),Young stellar objects (YSOs), X-ray emitting stars, High mass XRBs, Low-mass XRBs, Ultra-luminous X-ray sources (ULXs), Cataclysmic variables (CVs), and pulsars. \cite{kiker23} used a supervised learning method to detect and characterise the aperiodic variability in the XRB power spectra.

\begin{figure*}
    \centering
    \includegraphics[width=0.48\linewidth]{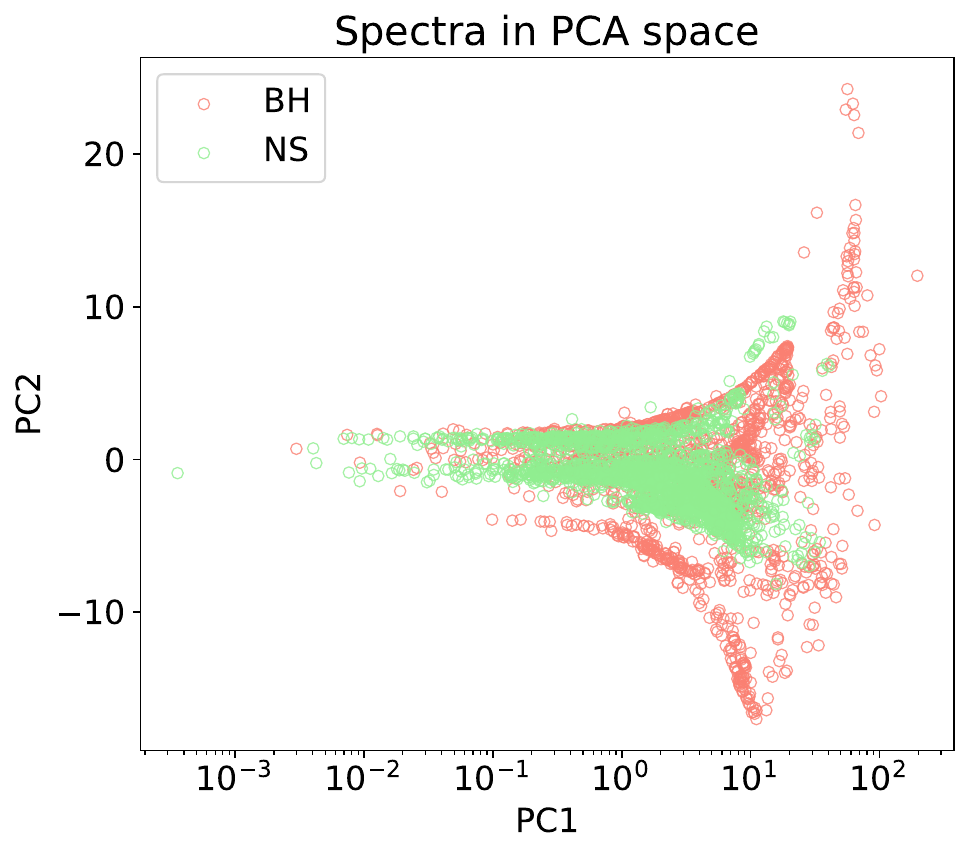}
    \includegraphics[width=0.48\linewidth]{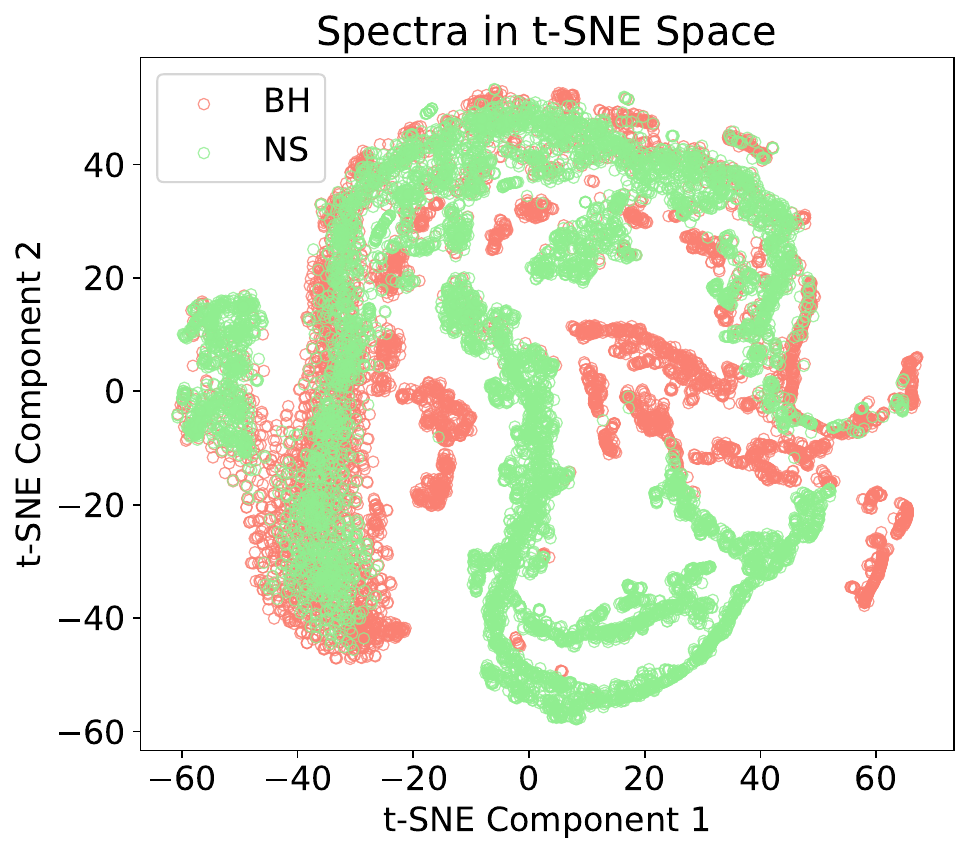}
    \caption{Two-dimensional projections of the full set of 43-dimensional RXTE energy spectra using linear and non-linear dimensionality-reduction techniques. The left panel shows the projection onto the first two principal components (PC1 and PC2) obtained via Principal Component Analysis (PCA), which preserves the directions of maximum variance in the original spectral space. The right panel shows the corresponding embedding produced by t-distributed Stochastic Neighbor Embedding (t-SNE), which emphasizes local neighborhood structure. Each point represents a single energy spectrum, with black hole X-ray binaries (BH XRBs) shown in red and neutron star X-ray binaries (NS XRBs) in green. These projections enable a qualitative assessment of overlap, clustering, and separability between the two source classes in reduced-dimensional space.}
    \label{fig:pca}
\end{figure*}

\cite{pattnaik} trained a random forest classifier on RXTE energy spectra (5–25 keV) of 61 sources to distinguish BH and NS XRB amongst them. The supervised algorithm achieved an average classification accuracy of $87 \pm 13\%$, and was utilised in the prediction of the nature of other low-mass XRBs of previously unknown type. This investigative study demonstrated that a classical machine learning technique, such as a random forest classifier, can reliably determine the class of a BH-XRB using only a 43-bin energy spectrum. It further showed that high signal-to-noise spectra in the high-soft state of XRBs can be classified with higher confidence. A feature-importance analysis revealed that both the lower- and higher-energy parts of the spectrum play a key role in distinguishing the classes.

However, classical machine learning methods, though simple, may miss important nonlinear correlations in the data. For example, \cite{pattnaik} showed that the random forest algorithm could separate the energy spectra of BH and NS XRBs by identifying the importance of specific energy bins, but this could not be directly linked to more interpretable physical characteristics of accretion disks, such as their radius or temperature, which are key to understanding the behavior of XRBs. It would be more useful if the XRB population could be separated based on their physical properties, placing strong constraints on accretion parameters and testing how well BH and NS systems can be distinguished. This would require neural networks, which can capture the nonlinearity in the data and, with the help of interpretable techniques, can reveal the accretion properties of XRBs that make them distinct. 

In light of this, we apply a simple Deep Learning Network to the RXTE dataset of XRB energy spectra, previously analyzed by \cite{pattnaik}, with the dual aim of improving the classification accuracy between BH and NS systems and identifying the physical properties that may drive the neural classifier.

In the next section, we describe the RXTE data used in this study. Section 3 presents the analysis with the DLN, where the energy spectrum is used as input, along with the obtained results. In Section 4, we discuss the empirical fitting of all energy spectra using traditional spectral fitting methods and how this motivated the development of a new DLN that directly uses the spectral parameters and their results. Finally, in the last section, we summarize and discuss the conclusions of this study.

\section{Data Selection} \label{sec:style}

\cite{pattnaik} carried out their classification analysis on 14,885 archival pointed RXTE/PCA observations of 61 low-mass XRBs, comprising 33 NS-XRBs and 28 BH-XRBs. They used the 5–25 keV energy spectra as input to their ML method, constructing for each observation an input vector of 43 channels by binning the time-averaged count rate spectrum. All spectra were response and background corrected before passing them to the ML network. In this work, we adopt the same dataset and energy spectra for training and testing our neural network classifier. Section 2 of \cite{pattnaik} provides further details on source selection, data reduction, and its preparation.

During our analysis, we found that RXTE observation IDs for one of the sources, named as XTEJ1908+094 in \cite{pattnaik}, actually correspond to a NS-XRB 4U 1907+097. We made this correction in our analysis, which now comprises 34 NS-XRBs and 27 BH-XRBs. Figure \ref{fig:observation} shows the reproduced version of Figure 1 in \cite{pattnaik} to give information about the total exposure (in ks) and average count rate (c/s)  within 5-25 keV, along with the total number of observations of each source. Figure \ref{fig:spectra} displays representative time-averaged energy spectra selected from observations with the highest average count rates for each of the BH- and NS-XRBs. It is evident that no distinct spectral features are visually apparent that could reliably distinguish the two types of systems. To explore whether more subtle differences exist in the spectral space, dimensionality reduction methods such as Principal Component Analysis (PCA) and t-distributed Stochastic Neighbor Embedding (t-SNE) were utilized to project the 43-dimensional energy spectra onto a two-dimensional latent feature space that preserves the dominant variance (PCA) and local neighborhood structure (t-SNE) of the original spectral distributions, enabling visual assessment of clustering and class separability. Figure \ref{fig:pca} shows the resulting PCA and t-SNE distributions, which reveal that the spectra do not separate easily into distinct groups. This motivates the use of neural network–based approaches capable of capturing complex nonlinear correlations in the spectral domain.

\begin{figure*}
    \centering
    \includegraphics[width=0.29\linewidth]{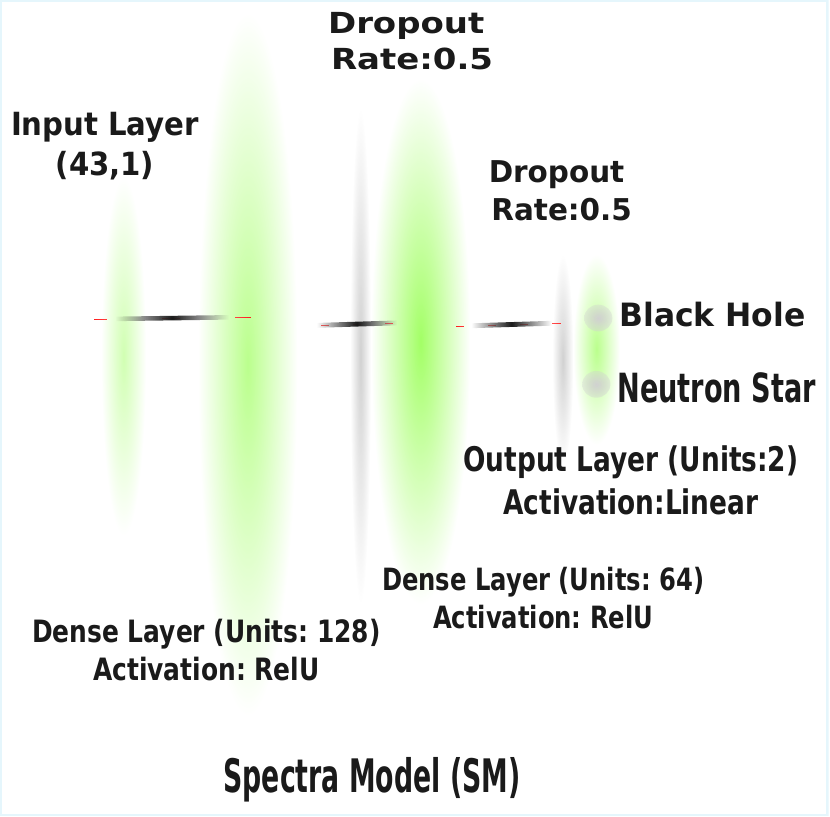}
;    \includegraphics[width=0.29\linewidth]{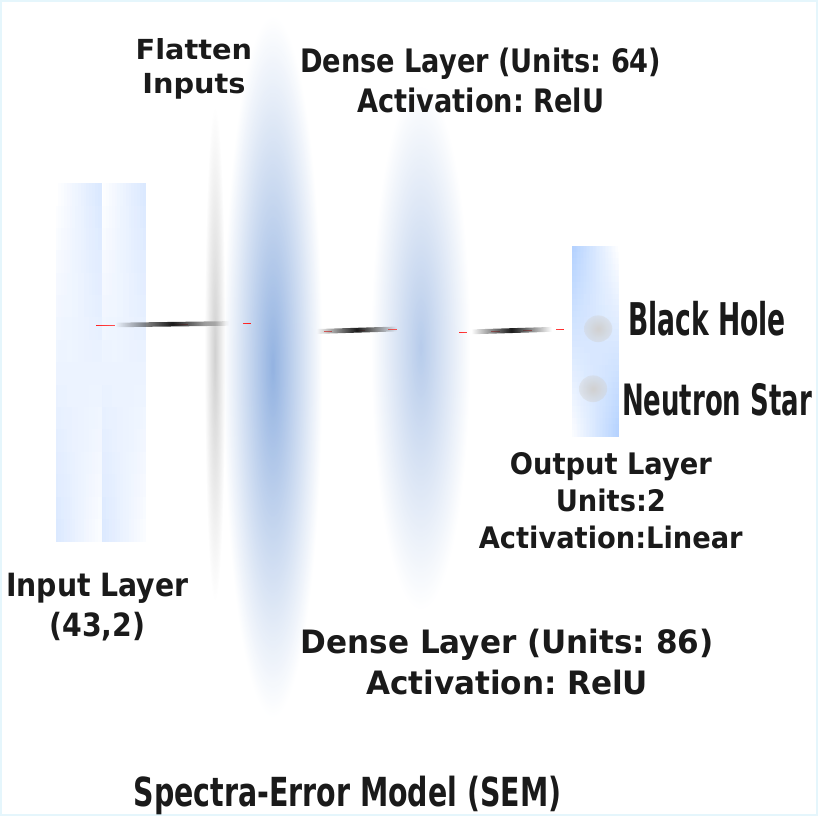}
    \includegraphics[width=0.29\linewidth]{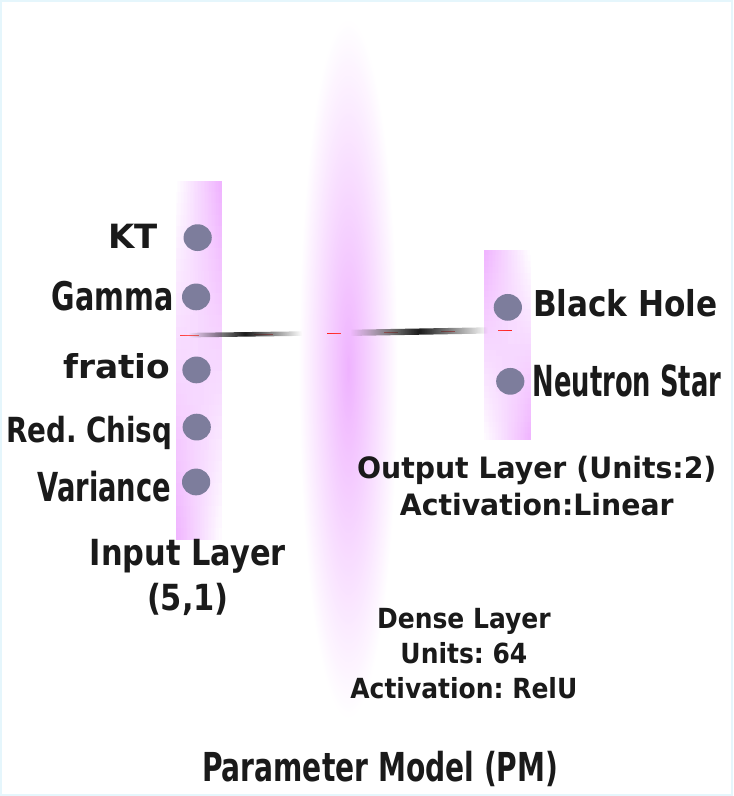}
    \caption{Neural network architectures used in this work. The left panel shows the Spectral Model (SM), which consists of a single input layer with 43 flux bins, followed by two hidden layers of different sizes and a two-node output layer. The middle panel displays the Spectral–Error Model (SEM), where the input layer is a two-dimensional array comprising 43 spectral flux bins and their corresponding errors, followed by two hidden layers and a two-node output layer. The right panel shows the Parameter Model (PM), which uses a five-dimensional input vector of best-fit spectral parameters: blackbody temperature (kT), photon index ($\Gamma$), blackbody-to-power-law flux ratio (fratio = F${bb}$/F${pl}$), reduced $\chi^2$, and variance, derived from spectral fitting, followed by a single hidden layer and a two-node output layer. The PM is discussed in section \ref{sec:highlight}. All colors used for the three networks are consistent with those in Figure \ref{fig:SM_SEM_scc}, which shows the accuracy distributions of the three models.}
    \label{fig:spectra-model}
\end{figure*}

\begin{figure}
    \centering
    \includegraphics[width=\linewidth]{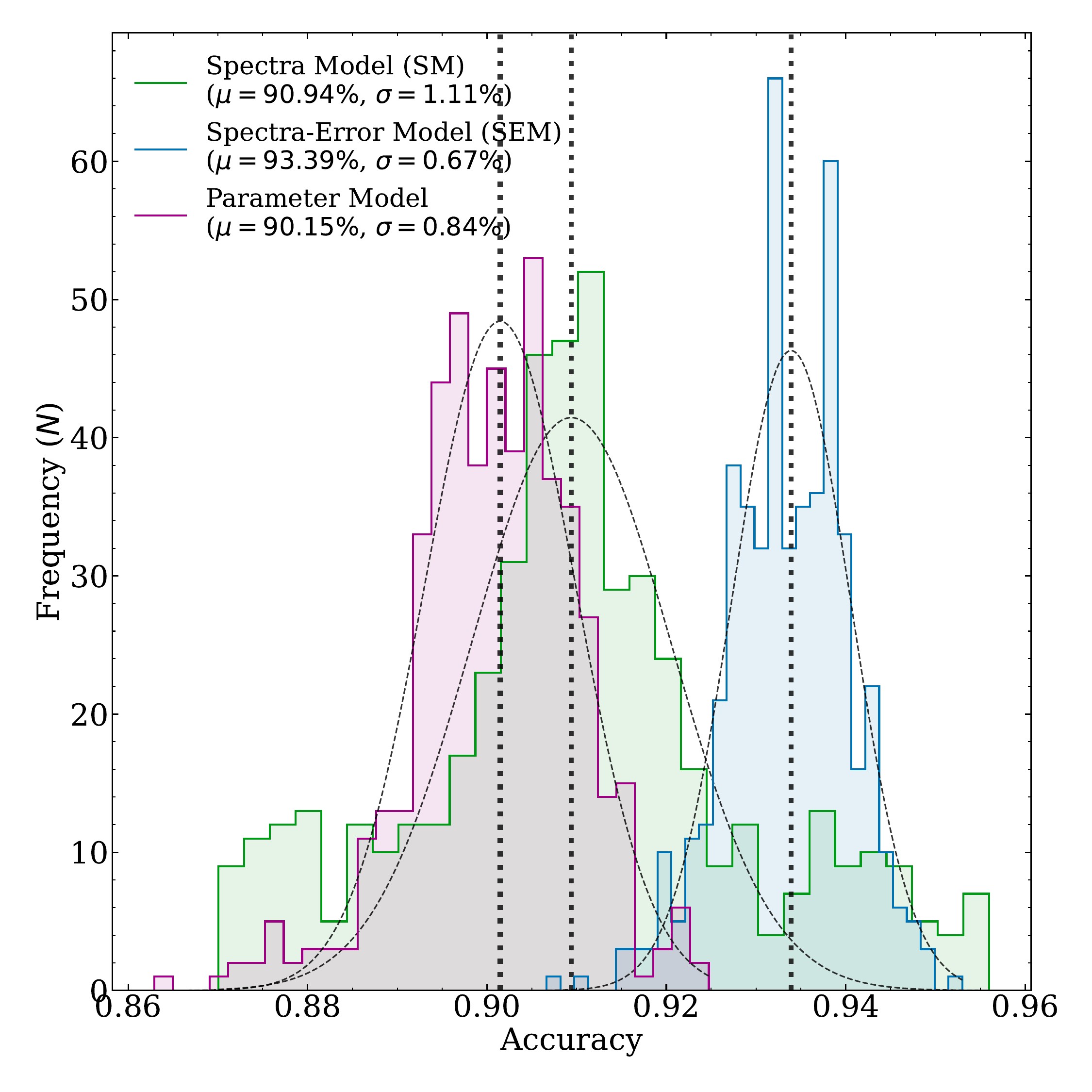}
    \caption{Comparison of accuracy distributions between the Spectra Model (SM, orange), Spectra-Error Model (SEM, blue), and the Parameter model (PM, green, see section \ref{sec:highlight}). Dashed lines represent Gaussian approximations of the underlying distributions, and vertical dotted lines mark the mean values. The SEM consistently outperforms the SM and PM, indicating the utility of incorporating error arrays into the training process.}
    \label{fig:SM_SEM_scc}
\end{figure}

\begin{figure*}[ht]
    \centering
    \includegraphics[width=0.9\linewidth]{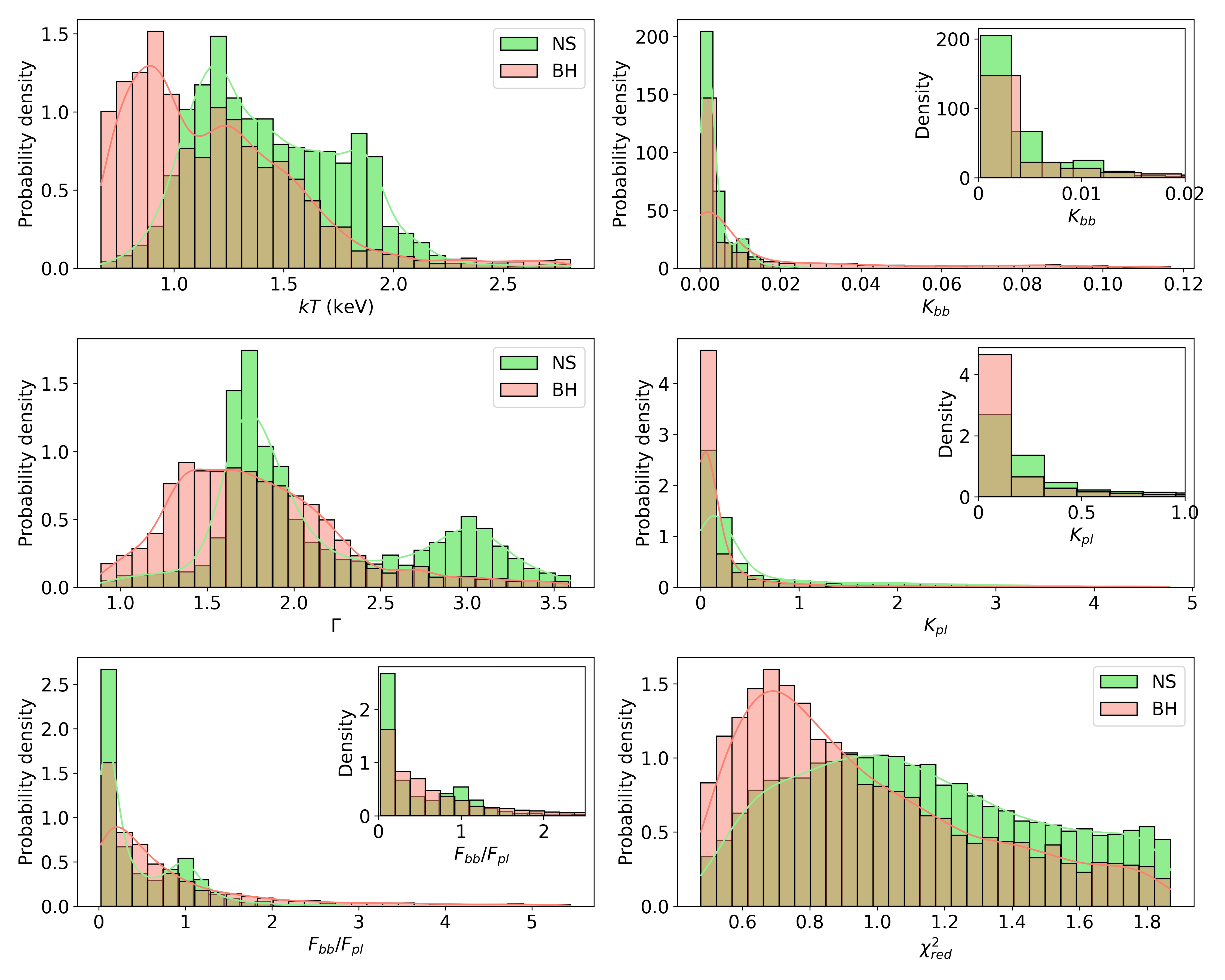}
    \caption{Normalized histograms of best-fit spectral parameters obtained from the spectral fitting, after applying a cut of $\chi^2_{\rm red} < 2$. The panels show the distributions of the blackbody temperature ($kT$), blackbody normalization, photon index ($\Gamma$), power-law normalization, blackbody-to-power-law flux ratio ($F_{\rm bb}/F_{\rm pl}$), and the reduced $\chi^2$. For each parameter, the distributions corresponding to the two classes (Class NS: blue; Class BH: red) are overlaid and normalized to unit area. To reduce the impact of extreme outliers, only values within the 5th–95th percentile range are shown. Inset panels highlight the low-value regions of the blackbody normalization, power-law normalization, and flux-ratio distributions.}
    \label{fig:parhisto}
\end{figure*}

\section{Deep Learning Network Classifier} \label{sec:floats}
We treat the Deep Learning Network (DLN) as a non-linear function approximator, $f: \mathbb{R}^n \to \mathbb{R}^m$. The network architecture is composed of interconnected units (neurons), where the output of a single neuron $j$ is given by $y_j = \sigma(\sum w_{ij}x_i + b_j)$. Here, $w_{ij}$ represents the weights, $b_j$ the bias, and $\sigma(\cdot)$ a non-linear activation function (e.g., ReLU or sigmoid) \citep{nair2010rectified}. This architecture allows the DLN to generalize patterns from training data. In the context of the Universal Approximation Theorem \citep{hornik1990universal, cybenko1989approximation}, such networks act as universal approximators, capable of mimicking the underlying physics or statistical distributions of astronomical datasets without explicit programming of the governing equations.
\\ \\
In this work, we employ a DLN as a spectral classifier for X-ray binaries, with the goal of distinguishing between BH and NS systems. The network learns directly from the spectra, mapping them to compact latent representations before outputting class probabilities. This approach leverages the DLN’s ability to capture non-linear dependencies in spectral data, enabling robust discrimination between the two classes of compact objects. 
\\ \\
We develop two different DLN models in this work:
\begin{enumerate}
    \item \textbf{Spectra Model (SM):} A DLN that takes in a spectrum as input and classifies it into BH or NS. 
    The input is a vector of size 43. The corresponding DLN architecture is shown in Fig.~\ref{fig:spectra-model}. 
    \item \textbf{Spectra-Error Model (SEM):} A DLN that takes in both the spectrum and the associated errors for each energy bin measurement. 
    The input is a vector of shape (43, 2) consisting of flux and errors. The corresponding architecture is depicted in Fig.~\ref{fig:spectra-model}.
\end{enumerate}

The rationale for constructing two distinct models is to investigate the role of measurement uncertainty in classification performance. While the SM relies solely on flux values across energy bins, the SEM explicitly incorporates the associated observational errors, allowing the DLN to learn how uncertainties shape the overall derived spectral features. By treating the flux–error pair jointly, SEM has the potential to improve robustness, particularly in low signal-to-noise regimes where classification is otherwise ambiguous.

Both models employ a feed-forward architecture with two hidden layers, nonlinear activation functions (ReLU for hidden layers, softmax for output), and dropout regularizers \citep{srivastava2014dropout} to mitigate overfitting as shown in Figure \ref{fig:spectra-model}. The networks are trained using categorical cross-entropy loss, which measures the dissimilarity between the predicted probability distribution $\hat{y}$ and the true one-hot encoded labels $y$. For a single sample with $C$ classes, it is defined as:
\begin{equation}
    L = - \sum ^ C _{c=1} y_c \log (\hat{y}_c)
\end{equation}
where $y_c$ is 1 if the sample belongs to class $c$ (and 0 otherwise), and $\hat{y}_c$ is the predicted probability for class $c$. Minimising this loss encourages the model to assign high probability to the correct class in other words, it penalizes the model when it assigns low confidence to the correct class.\\


\section{Statistical Robustness and Comparative Analysis}

\subsection{Monte Carlo Cross-Validation Strategy}
The training and evaluation of both the \textbf{SM} and the \textbf{SEM} were initially established using a standard 80:20 train-test split. All computational routines were executed on a workstation equipped with a single NVIDIA RTX 2060 GPU. However, relying on a single random partition of the dataset can introduce selection bias, where the reported accuracy is an artifact of a specific, favourable distribution of easy-to-classify objects in the test set. Furthermore, neural networks are stochastic by nature; their final performance can fluctuate based on the specific composition of the training data.

To rigorously assess the generalizability of our models and mitigate these selection biases, we employed a Monte Carlo cross-validation approach. Instead of a fixed validation set, we performed $N=500$ independent training and evaluation iterations. In each iteration, the dataset was randomly reshuffled and split, generating a unique combination of Observation IDs (ObsIDs) for training and testing. This exhaustive resampling ensures that the models are tested against a diverse range of spectral morphologies and signal-to-noise ratios, providing a statistical distribution of performance rather than a single point estimate.

Crucially, we enforced stratified sampling during these splits. Astronomical datasets are frequently imbalanced, often dominated by common sources while containing just a few examples of rare high-energy phenomena. Simple random splitting risks producing training or test sets that under-represent or entirely omit these rare classes, thereby biasing the learned decision boundaries and degrading generalisation performance. To mitigate this, we adopted a stratified train–test split, in which the data were partitioned conditional on the class labels so that each subset preserved the same class fractions as the full dataset. This ensured that both the training and testing sets contained representative examples of all spectral classes, including the rarest ones. The relative class proportions were thus maintained consistently across all 500 realisations, preserving the physical and statistical representativeness of the data.

\subsection{Performance Distribution and Error Integration}
Figure \ref{fig:SM_SEM_scc} presents the frequency distribution of accuracy scores derived from the 500 Monte Carlo iterations. This visualization offers a direct comparison of model stability and performance. The \textbf{SM} (orange histogram) exhibits a mean accuracy of approximately $90.94$ per cent, with a noticeably wider distribution ($\sigma = 1.11\%$). This variance suggests that the SM is more sensitive to the specific subset of data it is trained on, struggling to generalise well when it faces noisier or more complex spectra in the test set.

In contrast, the \textbf{SEM} (blue histogram) demonstrates a substantial improvement, achieving a mean accuracy of $\sim 93.39$ per cent. Perhaps more importantly, the SEM distribution is narrower and sharper ($\sigma = 0.67\%$). This reduction in variance indicates that the SEM is more statistically robust. By explicitly integrating observational error arrays into the input layer, the SEM gives lower importance to unreliable spectral bins (high noise) and greater importance to significant features. This capability effectively suppresses the influence of heteroscedastic noise, allowing the model to focus on the underlying physical signal. Consequently, the SEM yields higher accuracy with greater consistency across randomised data splits.

\section{Empirical spectral parameters based classification} \label{sec:highlight}

The time-averaged energy spectrum of an XRB is often fitted with a combination of a soft thermal component and a hard non-thermal radiative component. Typically, built-in models in the NASA spectral-fitting package XSPEC, such as {\sc bbody} and {\sc powerlaw}, are used to describe the low-energy blackbody disk emission and the high-energy coronal emission near the compact object, respectively. The {\sc bbody} model has two parameters: the disk temperature (kT, in keV) and a normalization parameter ($K_{bb}$). The {\sc powerlaw} model also has two parameters: the photon index ($\Gamma$) and a normalization parameter ($K_{pl}$).

To begin with, we fitted all the RXTE spectra of both BH and NS systems with the simple model combination {\sc bbody+powerlaw} to determine the spectral parameters and to examine whether BH and NS systems are distinguishable in the resulting parameter space of kT, $\Gamma$, $K_{bb}$, and $K_{pl}$. We found that this simple model was able to fit only 8,248 spectra with a reduced $\chi^2 < 2$. Figure \ref{fig:parhisto} shows the normalised histograms of the best-fit spectral parameters, the blackbody-to-power-law flux ratio (fratio = F$_{bb}$/F$_{pl}$), and the reduced $\chi^2$ for these accepted fits. A 3\% exclusion from both ends of each distribution is applied to remove extreme outliers and improve visual clarity. It is evident that the distributions for both BH and NS systems are largely overlapping for all parameters, making a sharp distinction between the two populations difficult. Nevertheless, this spectral fitting exercise indicates that the parameter vector comprising kT, $\Gamma$, fratio, and reduced $\chi^2$ provides a reasonable representation of the time-averaged spectrum. This motivated us to investigate whether such a 4-dimensional parameter vector, when given as input to the same DLNs described in Section \ref{sec:floats}, can yield a classification accuracy comparable to that obtained using the full 43-dimensional spectral data. If so, it would suggest that the two-layer DLNs are effectively relying on these parameters or their underlying distributions to perform the classification, thereby providing an astrophysical interpretation of the network’s decision process. 

Additionally, there is another important point to consider. The SM did not take any statistical errors in the spectra as input, yet it was able to classify the sources with high accuracy. This suggests that the network was implicitly learning the degree of variation present in the spectral shape across different cases. Such a measure of spectral variability could, in principle, be quantified and used as an additional parameter to complement the above 4-dimensional parameter vector. To explore this idea, we assigned a fiducial fractional error of 3\% to the flux values of each spectrum and refitted all spectra with the same {\sc bbody+powerlaw} model. The resulting reduced $\chi^2$ then provides a measure of how far the model deviates from the data relative to this arbitrarily chosen tolerance. Although this reduced $\chi^2$ carries no statistical meaning in the conventional sense, it serves as a proxy for the spectral variation that the SM might have been sensitive to. We therefore include this quantity, labelled ‘var’ or $\sigma^2$, as an additional parameter to form a 5-dimensional input vector, which we feed into the same neural network configuration. This will then become our third neural network model in this work, referred to as the \textbf{Parameter Model (PM}). It is a DLN that takes in a 5-d parameter vector as input and classifies it into BH or NS. The input is a vector of size 5, and its architecture is shown in Fig.~\ref{fig:spectra-model}.

\subsection{Quantifying Parameter Significance}
\begin{figure}
    \centering
    \includegraphics[width=\linewidth]{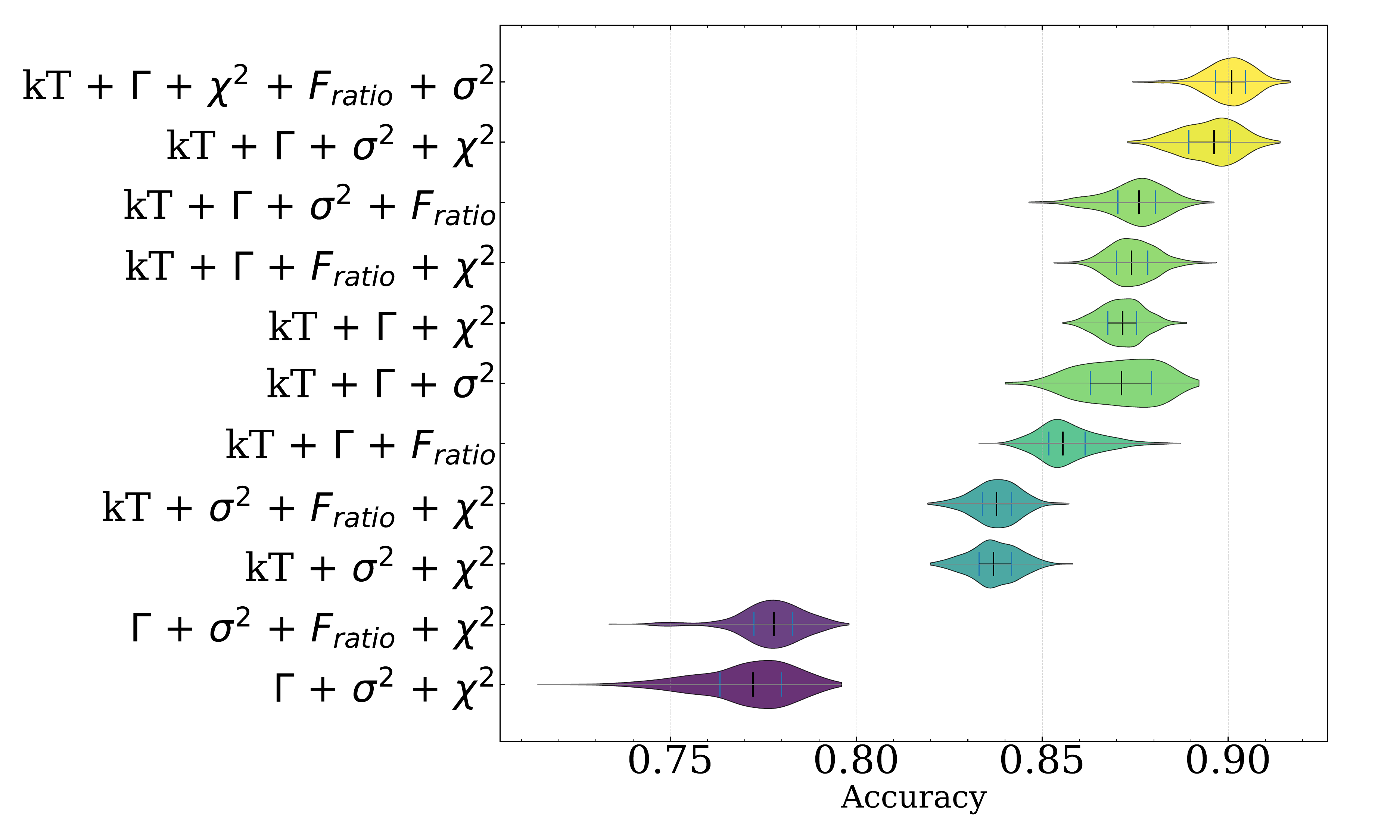}
    \caption{Impact of parameter selection on model performance. We compare the accuracy distributions for varying combinations of black body temperature ($kT$), photon index ($\Gamma$), reduced $\chi^2$, flux ratios ($F_{ratio}$), and variance ($\sigma^2$). The violin plots illustrate the density of accuracy scores across 500 randomised train-test splits. For each combination, the violin width represents the probability density of accuracies across split runs, while solid black markers denote the median accuracy, and blue markers enclose the middle 50 percent of the data (interquartile range). The results demonstrate a clear hierarchy, where the inclusion of auxiliary features ($\chi^2, \sigma^2$) alongside primary physical parameters significantly boosts both the median accuracy and the stability of the predictions.}
    \label{fig:feature_ablation}
\end{figure}

We extended the Monte Carlo methodology to the Parameter Model (\textbf{PM}) to assess not just performance, but also feature necessity. To avoid any biased analysis, we used the full dataset with all the spectra irrespective of the fit quality. We implemented a feature ablation analysis, where the model was retrained 500 times for each permutation of the input feature space. By systematically withholding parameters, such as the photon index ($\Gamma$) or variance ($\sigma^2$), we could isolate their marginal contribution to the classification accuracy. This approach allows us to distinguish between primary physical drivers and auxiliary statistical descriptors, providing a quantitative measure of feature importance under varying data splits, as illustrated in Figure \ref{fig:feature_ablation}. We found that models incorporating $kT$ and $\Gamma$ consistently outperform those relying on variability-only features, achieving median accuracies $\geq 85\%$. The model shows the best performance when all parameters are used ($kT + \Gamma + \chi^{2} + F_{\mathrm{ratio}} + \sigma^{2}$), with a median classification accuracy of $\simeq 90\%$ and a comparatively narrow spread. 
This indicates not only higher predictive power, but also a more stable and robust model across different realizations. In contrast, feature combinations that exclude the temperature parameter $kT$ consistently yield lower accuracies, typically in the range $\simeq 75$--$80 \%$, and exhibit broader distributions. This behavior underscores the central role of spectral temperature information in separating the classes. Overall, the figure illustrates that incorporating complementary spectral, variability, and goodness-of-fit features leads to the most reliable classification performance.

\section{Discussion and Conclusion} \label{sec:cite}

In X-ray astronomy, identifying the nature of the compact object in an unknown XRB system often requires detailed spectral and timing analysis. Previous works, such as \cite{pattnaik}, have addressed this classification issue using machine learning methods, including a random forest classifier, and demonstrated the efficacy of long-term archival RXTE observations in classifying between BH and NS XRBs using energy spectra with significant accuracy. However, there is a need to make such machine learning methods interpretable and gain astrophysical insights. In this work, we have addressed the interpretability issue using deep-learning–based classifier models that achieve comparable or improved classification accuracy while also providing insights into the astrophysical patterns that may be used by neural networks to perform the classification.

The general exploration of the X-ray spectra and their further reduction in PCA and t-SNE spaces does not reveal any clear distinction between the two classes. However, it does confirm the non-linearity of the problem at hand. As a first step, when we fed the normalised spectral arrays of all observations to the SM network, it achieved a classification accuracy of $\sim$90\%. The standard feature-importance test revealed that the network primarily uses both the high-energy and low-energy features of the spectra to arrive at its decision. However, a bias may arise for the high-energy half due to larger uncertainties in each spectral flux value, which are not included in the SM network. Therefore, when we passed the errors alongside the flux values of the spectral arrays to the SEM network, the classification accuracy improved to $\sim$94\%. The feature-importance test again showed that low-energy features remain important even after accounting for the errors. Although these results provide some indication that the soft energy component, believed to originate from regions farther away from the compact object, is important, the role of physical accretion parameters, such as disc temperature and radius, remains unclear.

Typically, the X-ray spectra of BH and NS binaries are described using a thermal disk component and a non-thermal component. Accordingly, we fitted all the spectra with a simple model consisting of blackbody emission from the disk and a power-law component originating from the inner regions around the compact object. From these models, we obtained parameters such as the disc temperature and the power-law photon index, which are known to be correlated with the electron temperature and optical depth of the hot corona, along with the fluxes of both components. However, it is well known that such a simple model does not fully describe X-ray spectra and that additional emission and absorption features are often required. Therefore, instead of increasing the model complexity, we included the reduced $\chi^2$ and the variance of the data to account for model inadequacies. The histograms of the fitted parameters, kT and $\Gamma$, show that NS sources tend to occupy higher temperature regions and exhibit softer spectra compared to BH sources. However, there is substantial overlap in the ranges  0.9$<$kT$<$2 keV and 1.1$<\Gamma<$2.7, which makes it difficult to distinguish the two populations using traditional methods. Nevertheless, this suggests that these fitted parameters, along with additional unaccounted features, may be the quantities that the neural network exploits to arrive at its classification.

Using the PM network, we find that a compact five-parameter space can also achieve a classification accuracy comparable to that obtained using the full spectral information. Figure \ref{fig:SM_SEM_scc} shows that all three models employed in this work yield similar accuracies, indicating that the neural network, when trained on spectra, is effectively encoding information related to the disc temperature, power-law index, the ratio of disc to power-law flux, the reduced $\chi^2$, and the variance of the data. This is an important result, as it suggests that the distinction between BH and NS systems can be made using physically meaningful parameters rather than relying on satellite-specific, response-dependent spectral shapes. We further investigated the relative importance of these five parameters by systematically removing one parameter at a time from the five-dimensional input space and re-evaluating the classification accuracy. This analysis shows that kT and $\Gamma$ contribute most strongly to the classification performance. Both of these parameters describe the accretion phenomenon in BH and NS sources, and it is therefore encouraging to see that the machine-learning network relies on the same physically motivated parameters to perform the classification.


\begin{acknowledgments}
We would like to express our gratitude to the High Energy Astrophysics Science Archive Research Centre (HEASARC) for providing the software and data utilised in this project. We are grateful to the authors of \cite{pattnaik} for their detailed work on the RXTE data set and for providing access to the data used in our analysis.
\end{acknowledgments}

\bibliography{sample701}{}

@ARTICLE{orwat22,
       author = {{Orwat-Kapola}, Jakub K. and {Bird}, Antony J. and {Hill}, Adam B. and {Altamirano}, Diego and {Huppenkothen}, Daniela},
        title = "{Light-curve fingerprints: an automated approach to the extraction of X-ray variability patterns with feature aggregation - an example application to GRS 1915+105}",
      journal = {\mnras},
         year = 2022,
        month = jan,
       volume = {509},
       number = {1},
        pages = {1269-1290},
          doi = {10.1093/mnras/stab3043},
archivePrefix = {arXiv},
       eprint = {2110.10063},
 primaryClass = {astro-ph.IM},
       adsurl = {https://ui.adsabs.harvard.edu/abs/2022MNRAS.509.1269O}
}

@ARTICLE{kiker23,
       author = {{Kiker}, Thaddaeus J. and {Steiner}, James F. and {Garraffo}, Cecilia and {M{\'e}ndez}, Mariano and {Zhang}, Liang},
        title = "{QPOML: a machine learning approach to detect and characterize quasi-periodic oscillations in X-ray binaries}",
      journal = {\mnras},
         year = 2023,
        month = oct,
       volume = {524},
       number = {4},
        pages = {4801-4818},
          doi = {10.1093/mnras/stad1643},
archivePrefix = {arXiv},
       eprint = {2306.04055},
 primaryClass = {astro-ph.HE},
       adsurl = {https://ui.adsabs.harvard.edu/abs/2023MNRAS.524.4801K}
}

@ARTICLE{beurs22,
       author = {{de Beurs}, Zoe L. and {Islam}, N. and {Gopalan}, G. and {Vrtilek}, S.~D.},
        title = "{A Comparative Study of Machine-learning Methods for X-Ray Binary Classification}",
      journal = {\apj},
         year = 2022,
        month = jul,
       volume = {933},
       number = {1},
          eid = {116},
        pages = {116},
          doi = {10.3847/1538-4357/ac6184},
archivePrefix = {arXiv},
       eprint = {2204.00346},
 primaryClass = {astro-ph.HE},
       adsurl = {https://ui.adsabs.harvard.edu/abs/2022ApJ...933..116D}
}

@ARTICLE{kumaran23,
       author = {{Kumaran}, Shivam and {Mandal}, Samir and {Bhattacharyya}, Sudip and {Mishra}, Deepak},
        title = "{Automated classification of Chandra X-ray point sources using machine learning methods}",
      journal = {\mnras},
         year = 2023,
        month = apr,
       volume = {520},
       number = {4},
        pages = {5065-5076},
          doi = {10.1093/mnras/stad414},
archivePrefix = {arXiv},
       eprint = {2302.09008},
 primaryClass = {astro-ph.HE},
       adsurl = {https://ui.adsabs.harvard.edu/abs/2023MNRAS.520.5065K}
}

@ARTICLE{daniela17,
       author = {{Huppenkothen}, Daniela and {Heil}, Lucy M. and {Hogg}, David W. and {Mueller}, Andreas},
        title = "{Using machine learning to explore the long-term evolution of GRS 1915+105}",
      journal = {\mnras},
         year = 2017,
        month = apr,
       volume = {466},
       number = {2},
        pages = {2364-2377},
          doi = {10.1093/mnras/stw3190},
archivePrefix = {arXiv},
       eprint = {1611.01332},
 primaryClass = {astro-ph.HE},
       adsurl = {https://ui.adsabs.harvard.edu/abs/2017MNRAS.466.2364H}
}

@ARTICLE{pattnaik,
       author = {{Pattnaik}, R. and {Sharma}, K. and {Alabarta}, K. and {Altamirano}, D. and {Chakraborty}, M. and {Kembhavi}, A. and {M{\'e}ndez}, M. and {Orwat-Kapola}, J.~K.},
        title = "{A machine-learning approach for classifying low-mass X-ray binaries based on their compact object nature}",
      journal = {\mnras},
         year = 2021,
        month = mar,
       volume = {501},
       number = {3},
        pages = {3457-3471},
          doi = {10.1093/mnras/staa3899},
archivePrefix = {arXiv},
       eprint = {2012.06934},
 primaryClass = {astro-ph.HE},
       adsurl = {https://ui.adsabs.harvard.edu/abs/2021MNRAS.501.3457P}
}

@BOOK{frank85,
       author = {{Frank}, J. and {King}, A.~R. and {Raine}, D.~J.},
        title = "{Accretion power in astrophysics}",
         year = 1985,
       adsurl = {https://ui.adsabs.harvard.edu/abs/1985apa..book.....F}
}

@INCOLLECTION{belloni10,
       author = {{Belloni}, T.~M.},
        title = "{States and Transitions in Black Hole Binaries}",
    booktitle = {Lecture Notes in Physics, Berlin Springer Verlag},
         year = 2010,
       editor = {{Belloni}, Tomaso},
       volume = {794},
        pages = {53},
          doi = {10.1007/978-3-540-76937-8_3},
       adsurl = {https://ui.adsabs.harvard.edu/abs/2010LNP...794...53B}
}

@ARTICLE{liu07,
       author = {{Liu}, Q.~Z. and {van Paradijs}, J. and {van den Heuvel}, E.~P.~J.},
        title = "{A catalogue of low-mass X-ray binaries in the Galaxy, LMC, and SMC (Fourth edition)}",
      journal = {\aap},
         year = 2007,
        month = jul,
       volume = {469},
       number = {2},
        pages = {807-810},
          doi = {10.1051/0004-6361:20077303},
archivePrefix = {arXiv},
       eprint = {0707.0544},
 primaryClass = {astro-ph},
       adsurl = {https://ui.adsabs.harvard.edu/abs/2007A&A...469..807L}
}

@ARTICLE{avakyan23,
       author = {{Avakyan}, A. and {Neumann}, M. and {Zainab}, A. and {Doroshenko}, V. and {Wilms}, J. and {Santangelo}, A.},
        title = "{XRBcats: Galactic low-mass X-ray binary catalogue}",
      journal = {\aap},
         year = 2023,
        month = jul,
       volume = {675},
          eid = {A199},
        pages = {A199},
          doi = {10.1051/0004-6361/202346522},
archivePrefix = {arXiv},
       eprint = {2303.16168},
 primaryClass = {astro-ph.HE},
       adsurl = {https://ui.adsabs.harvard.edu/abs/2023A&A...675A.199A}
}

@ARTICLE{fortin24,
       author = {{Fortin}, F. and {Kalsi}, A. and {Garc{\'\i}a}, F. and {Simaz-Bunzel}, A. and {Chaty}, S.},
        title = "{A catalogue of low-mass X-ray binaries in the Galaxy: From the INTEGRAL to the Gaia era}",
      journal = {\aap},
         year = 2024,
        month = apr,
       volume = {684},
          eid = {A124},
        pages = {A124},
          doi = {10.1051/0004-6361/202347908},
archivePrefix = {arXiv},
       eprint = {2401.11931},
 primaryClass = {astro-ph.HE},
       adsurl = {https://ui.adsabs.harvard.edu/abs/2024A&A...684A.124F}
}

@ARTICLE{santana16,
       author = {{Corral-Santana}, J.~M. and {Casares}, J. and {Mu{\~n}oz-Darias}, T. and {Bauer}, F.~E. and {Mart{\'\i}nez-Pais}, I.~G. and {Russell}, D.~M.},
        title = "{BlackCAT: A catalogue of stellar-mass black holes in X-ray transients}",
      journal = {\aap},
         year = 2016,
        month = mar,
       volume = {587},
          eid = {A61},
        pages = {A61},
          doi = {10.1051/0004-6361/201527130},
archivePrefix = {arXiv},
       eprint = {1510.08869},
 primaryClass = {astro-ph.HE},
       adsurl = {https://ui.adsabs.harvard.edu/abs/2016A&A...587A..61C}
}

@INCOLLECTION{bahramian23,
       author = {{Bahramian}, Arash and {Degenaar}, Nathalie},
        title = "{Low-Mass X-ray Binaries}",
    booktitle = {Handbook of X-ray and Gamma-ray Astrophysics},
         year = 2023,
          eid = {120},
        pages = {120},
          doi = {10.1007/978-981-16-4544-0_94-1},
       adsurl = {https://ui.adsabs.harvard.edu/abs/2023hxga.book..120B}
}

@ARTICLE{belloni18,
       author = {{Belloni}, Tomaso M.},
        title = "{X-ray emission from black-hole and neutron-star binaries}",
      journal = {arXiv e-prints},
         year = 2018,
        month = mar,
          eid = {arXiv:1803.03641},
        pages = {arXiv:1803.03641},
          doi = {10.48550/arXiv.1803.03641},
archivePrefix = {arXiv},
       eprint = {1803.03641},
 primaryClass = {astro-ph.HE},
       adsurl = {https://ui.adsabs.harvard.edu/abs/2018arXiv180303641B}
}

@INCOLLECTION{gilfanov10,
       author = {{Gilfanov}, M.},
        title = "{X-Ray Emission from Black-Hole Binaries}",
    booktitle = {Lecture Notes in Physics, Berlin Springer Verlag},
         year = 2010,
       editor = {{Belloni}, Tomaso},
       volume = {794},
        pages = {17},
          doi = {10.1007/978-3-540-76937-8_2},
       adsurl = {https://ui.adsabs.harvard.edu/abs/2010LNP...794...17G}
}

@ARTICLE{lecun15,
       author = {{LeCun}, Yann and {Bengio}, Yoshua and {Hinton}, Geoffrey},
        title = "{Deep learning}",
      journal = {\nat},
         year = 2015,
        month = may,
       volume = {521},
       number = {7553},
        pages = {436-444},
          doi = {10.1038/nature14539},
       adsurl = {https://ui.adsabs.harvard.edu/abs/2015Natur.521..436L}
}

@book{mitchell1997,
  title={Machine learning},
  author={Mitchell, Tom M},
  year={1997},
  publisher={McGraw-Hill},
  address={New York}
}

@article{hornik1990universal,
  title={Universal approximation of an unknown mapping and its derivatives using multilayer feedforward networks},
  author={Hornik, Kurt and Stinchcombe, Maxwell and White, Halbert},
  journal={Neural networks},
  volume={3},
  number={5},
  pages={551--560},
  year={1990},
  publisher={Elsevier}
}

@article{kembhavi2022machine,
  title={Machine learning in astronomy},
  author={Kembhavi, Ajit and Pattnaik, Rohan},
  journal={Journal of Astrophysics and Astronomy},
  volume={43},
  number={2},
  pages={76},
  year={2022},
  publisher={Springer}
}

@article{sen2022astronomical,
  title={Astronomical big data processing using machine learning: A comprehensive review},
  author={Sen, Snigdha and Agarwal, Sonali and Chakraborty, Pavan and Singh, Krishna Pratap},
  journal={Experimental Astronomy},
  volume={53},
  number={1},
  pages={1--43},
  year={2022},
  publisher={Springer}
}

@article{baron2019machine,
  title={Machine learning in astronomy: A practical overview},
  author={Baron, Dalya},
  journal={arXiv preprint arXiv:1904.07248},
  year={2019}
}

@article{barchi2020machine,
  title={Machine and Deep Learning applied to galaxy morphology-A comparative study},
  author={Barchi, Paulo H and de Carvalho, RR and Rosa, Reinaldo R and Sautter, RA and Soares-Santos, Marcelle and Marques, Bruno AD and Clua, Esteban and Gon{\c{c}}alves, TS and de S{\'a}-Freitas, C and Moura, TC},
  journal={Astronomy and Computing},
  volume={30},
  pages={100334},
  year={2020},
  publisher={Elsevier}
}

@article{cavuoti2015photometric,
  title={Photometric redshift estimation based on data mining with PhotoRApToR},
  author={Cavuoti, Stefano and Brescia, Massimo and De Stefano, Virgilio and Longo, Giuseppe},
  journal={Experimental Astronomy},
  volume={39},
  number={1},
  pages={45--71},
  year={2015},
  publisher={Springer}
}

@inproceedings{nair2010rectified,
  title={Rectified linear units improve restricted boltzmann machines},
  author={Nair, Vinod and Hinton, Geoffrey E},
  booktitle={Proceedings of the 27th international conference on machine learning (ICML-10)},
  pages={807--814},
  year={2010}
}

@article{cybenko1989approximation,
  title={Approximation by superpositions of a sigmoidal function},
  author={Cybenko, George},
  journal={Mathematics of control, signals and systems},
  volume={2},
  number={4},
  pages={303--314},
  year={1989},
  publisher={Springer}
}

@article{srivastava2014dropout,
  title={Dropout: a simple way to prevent neural networks from overfitting},
  author={Srivastava, Nitish and Hinton, Geoffrey and Krizhevsky, Alex and Sutskever, Ilya and Salakhutdinov, Ruslan},
  journal={The journal of machine learning research},
  volume={15},
  number={1},
  pages={1929--1958},
  year={2014},
  publisher={JMLR. org}
}

@article{bhatta2024gamma,
  title={Gamma-ray blazar classification using machine learning with advanced weight initialization and self-supervised learning techniques},
  author={Bhatta, Gopal and Gharat, Sarvesh and Borthakur, Abhimanyu and Kumar, Aman},
  journal={Monthly Notices of the Royal Astronomical Society},
  volume={528},
  number={1},
  pages={976--986},
  year={2024},
  publisher={Oxford University Press}
}

@article{hayat2021self,
  title={Self-supervised representation learning for astronomical images},
  author={Hayat, Md Abul and Stein, George and Harrington, Peter and Luki{\'c}, Zarija and Mustafa, Mustafa},
  journal={The Astrophysical Journal Letters},
  volume={911},
  number={2},
  pages={L33},
  year={2021},
  publisher={IOP Publishing}
}

@article{huertas2023brief,
  title={A brief review of contrastive learning applied to astrophysics},
  author={Huertas-Company, Marc and Sarmiento, Regina and Knapen, Johan H},
  journal={RAS Techniques and Instruments},
  volume={2},
  number={1},
  pages={441--452},
  year={2023},
  publisher={Oxford University Press}
}
\bibliographystyle{aasjournalv7}



\end{document}